\documentclass[%
 reprint,
showpacs,preprintnumbers,
nofootinbib,
 amsmath,amssymb,
 aps,
 twocolumn,
]{revtex4}

\usepackage{graphicx}% Include figure files
\usepackage{dcolumn}% Align table columns on decimal point
\usepackage{bm}% bold math
\newcommand{\CO}{{\cal O}}

\begin{document}

\title{Affleck-Dine baryogenesis with modulated reheating}

\author{Kohei Kamada}%
\email{kamada@resceu.s.u-tokyo.ac.jp}
\affiliation{Department of Physics, Graduate School of Science,
The University of Tokyo, Tokyo 113-0033, Japan}
\affiliation{Research Center for the Early Universe (RESCEU),
Graduate School of Science, The University of Tokyo, Tokyo 113-0033, Japan}

\author{Kazunori Kohri}
\email{kohri@post.kek.jp}
\affiliation{Cosmophysics group, Theory Center, Institute of Particle and Nuclear Studies, KEK,
1-1 Oho, Tsukuba 305-0801, Japan}

\author{Shuichiro Yokoyama}
 \email{shu@a.phys.nagoya-u.ac.jp}
\affiliation{%
Department of Physics and Astrophysics, Nagoya University, Aichi, 466-8602, Japan
}%

\date{\today}
\begin{abstract}

 Modulated reheating scenario is one of the most attractive models
that predict possible 
detections of not only the primordial non-Gaussianity but also 
the tensor fluctuation through future CMB observations such as the
Planck satellite, the PolarBeaR and the LiteBIRD satellite
experiments. 
  We study the baryonic-isocurvature fluctuations in
the Affleck-Dine baryogenesis with the modulated reheating
scenario.  We show that the Affleck-Dine baryogenesis can be
consistent with the modulated reheating scenario with respect to
the current observational constraint on the baryonic-isocurvature
fluctuations.

\end{abstract}

\preprint{KEK-TH-1385, KEK-Cosmo-42, RESCEU-19/10}

\pacs{98.80.Cq, 98.80.Es}

\maketitle

%\tableofcontents

\section{Introduction}

The observed cosmological perturbation is generated from a primordial
curvature perturbation. It is known that the curvature perturbation
can be time-independent after the cosmic time approximately  becomes
one second ($t \gtrsim 1$ sec)  because the cosmic fluid had been
perfectly radiation-dominated  until that time~\cite{RDcond}. We
expect the corresponding perturbations can be generated from the
vacuum fluctuation of a light field, which requires the mass of the
field to be lighter than the Hubble parameter $H$ during
inflation~\cite{textbook}. In original single-field slow-roll
inflation paradigm, the curvature perturbation mainly originated  from
the perturbation of the so-called inflaton field which induces the primordial
inflation. Then the curvature perturbation had been already produced
at the initial epoch during inflation and is constant. In this case, a
deviation from the Gaussian statistics of the fluctuation (so called
non-Gaussianity) is small since the perturbation  does not change the
relation between the energy density $\rho$ and the pressure density
$P$ very much through the cosmic history, and the non-linearity
parameter to express the degree of the non-Gaussianity should be  the
order of the slow-roll parameters $f_{\rm NL} \sim \epsilon~{\rm
and/or}~\eta \sim {\cal O}(10^{-2})$. Then it means that we will not
be able to detect the non-Gaussianity.

When we consider other paradigms by assuming other light field
$\sigma$, the curvature
perturbation may originate mainly  from the perturbation of
$\sigma$. Although the curvature perturbation from $\delta \sigma$
might be initially negligible, it will grow later even after inflation
ends  because it generates a non-adiabatic pressure
perturbation. This mechanism works at any epochs except for complete
radiation domination or  complete matter domination. In this case we
should be able to detect those different types of non-Gaussianity in
future.

So far a variety of  models  which produce a non-adiabatic pressure
perturbation and generate non-Gaussianity even in terms of slow-roll
inflation have been reported in this direction; during multi-field
inflation \cite{starobinsky:1985,Alabidi:2006hg}, before a second
reheating through the curvaton mechanism~\cite{curvatonNG}, at the end
of inflation
\cite{Bernardeau:2003,Lyth:2005,Huang:2009vk,Alabidi:2006wa,Lin:2009vj},
during modulated
reheating~\cite{modureh,Mazumdar:2003va,Ackerman:2005,Podolsky:2006,Suyama:2007}
or modulated preheating~\cite{Kohri:2009ac,otherModpre}, at a
modulated phase transition~\cite{Matsuda:2009yt,Kawasaki:2009hp}, and
a modulated trapping~\cite{Langlois:2009jp}. By observing the
non-Gaussianity, we will be able to discern a better model from other
ones. In this paper, especially we focus on modulated reheating
because of some attractive reasons mentioned later.

The latest constraint on the non-linearity parameter $f_{\rm NL}$ from
the WMAP 7-year data is $f_{\rm NL} = 32 \pm 21~(68\% {\rm
C.L.})$~\cite{Komatsu:2010fb} for the local type of non-Gaussianity,
which  means a null detection of the non-Gaussianity. On the other
hand, the Planck satellite is expected to reach $\Delta f_{\rm NL}\sim$
5.~\cite{PLANCK_HP,PLANCK_coll,Komatsu:2009kd}. Therefore we may be able to
detect the non-Gaussianity in near future within 5 years.

Another good indicators to discern a  model from others must be  the
tensor to scalar ratio $r$ for the perturbation. So far WMAP 7-year
has reported only its upper bound $r \lesssim
0.2$.~\cite{Komatsu:2010fb}, which does not exclude even the standard
single-field quadratic chaotic inflation model. It  is expected  that
the PLANCK satellite will be  able to  observe $r \sim
0.1$~\cite{PLANCK_HP,PLANCK_coll}. In addition, it should be quite
exciting that future ground-based detectors QUIET+PolarBeaR will reach
$r \sim {\cal O}(0.01)$~\cite{Hazumi:2008zz}, and  KEK's future CMB
satellite experiment, LiteBIRD will observe   $r  \sim {\cal
O}(10^{-3})$.~\cite{Hazumi:2008zz,LiteBird}

Considering these current and future situations for the detectability
of non-Gaussianity and the tensor fluctuation,  for comparison in
future it should be necessary to keep those known models in proper
order with respect to the  predicted values  of  $r$ and $f_{\rm NL}$,
respectively. However, unfortunately it might be known that there
exist few models which predict  both  the large non-Gaussianity of
$f_{\rm NL} \sim {\cal O}(10)$ and the large tensor to scalar ratio of
$r \sim {\cal O} (0.1)$. Therefore, it should be required for
theorists in advance to build models which predict both the large
$f_{\rm NL}$ and the large $r$  enough to be detectable in near future.

For this purpose, the modulated reheating model must be attractive
because it possibly  produces large non-Gaussianity and can predict
a large tensor to scalar ratio if the inflation scale is sufficiently large  like
chaotic inflation with its Hubble parameter $H \sim 10^{13}$ GeV.

On the other hand,  we also  have  to  check consistencies of the
modulated reheating scenario with other observational  constraints. As
will  be discussed later in detail, in the modulated reheating
scenario the curvature perturbation is effectively governed by  the
fluctuation of  the reheating temperature  after inflation $\delta
T_{R}/T_{R}$.  Since   most class  of viable baryogenesis scenarios in
modern cosmology depend on the reheating temperature, the modulated
reheating may induce a large baryonic-isocurvature
fluctuation.~\footnote{Some class of scenarios for dark-matter
production also depend on the reheating temperature such as gravitino
thermal/non-thermal production. In this case the modulated reheating is
severely constrained by observations of the cold dark matter
(CDM)-isocurvature fluctuation~\cite{Takahashi:2009dr,Takahashi:2009cx}. 
In this paper we are assuming a dark matter such as the lightest neutralino which
was decoupled from the thermal bath with its appropriate thermal-relic
density to fit the observation. Then the modulated reheating does not
produce a sizable CDM-isocurvature fluctuation.}  In this paper we
consider baryogenesis in models with supersymmetric (SUSY) extension
of standard model, especially so-called Affleck-Dine (AD)
mechanism~\cite{Affleck:1984fy,Dine:1995kz}, which is naturally realized  even in
the Minimal Supersymmetric Standard Model (MSSM)~\footnote{See also
Ref.~\cite{Mazumdar:2003va} for another mechanism of baryogenesis in
the SUSY cosmologies and its compatibility with the modulated
scenario.}  and agrees with observations in broad parameter regions.
Since  good candidates for the light scalar field $\sigma$ could be found in
SUSY~\cite{modureh} or supergravity (the local theory of SUSY), this
direction of discussion should be naturally motivated. As we have
already raised the question, however, it might be nontrivial if the
AD baryogenesis is consistent with the modulated reheating
scenarios because the produced baryon number sometimes depends on the
reheating temperature  in the normal parameter space. Thus we have a
strong motivation to search the allowed parameter region for the 
AD baryogenesis  to avoid the constraint on the
baryonic-isocurvature fluctuation  in the modulated reheating scenario.

This paper is organized as follows. In Sec. II we show the basic picture
of the modulated reheating scenarios and the conditions for the decay
rate of the inflaton field where the large non-Gaussianity can be
predicted. The AD baryogenesis is outlined in Sec.~III.  In
Sec. IV we look for the parameter space to agree with the observations and
also discuss a possible isocurvature fluctuation that the AD
baryogenesis may originally have. Sec. V is devoted to conclusions.

\section{Modulated reheating scenario}

Here, we give a brief review of the modulated reheating scenario~\cite{modureh}.
In such scenario, 
we consider 
the decay rate of the inflaton, $\Gamma$, depending on a light scalar field, $\sigma$,
which has a quantum fluctuation during inflation, that is,
$\Gamma = \Gamma (\sigma)$.
The $e$-folding number $N = \int d \ln a $, where $a$ is a scale factor, measured between the end of inflation at $t=t_{\rm inf}$ and
a time after the end of the complete reheating, $t_{\rm c}$, is given by
\begin{eqnarray}
N &\! = \!& \ln \left( {a (t_{\rm c}) \over a (t_{\rm inf})}\right) \nonumber\\
&\! = \!& \ln \left( {a (t_{\rm reh}) \over a (t_{\rm inf})} \right) 
+ \ln \left( {a(t_{\rm c}) \over a(t_{\rm reh}) } \right)~,
\label{eq:deltaN1}
\end{eqnarray}
where $t_{\rm reh}$ represents a time at $d\ln a /d t = H = \Gamma$. 

Let us consider the quadratic inflaton potential, $V(\phi) \propto \phi^2$.
In such case, during the inflaton oscillating phase after the inflation,
the energy density of the Universe relying on the inflaton decays as $\rho \propto a^{-3}$ 
and the Hubble parameter, $H$, evolves as
$H \propto \rho^{1/2}$. 
Since after the complete reheating the energy density of the Universe is 
dominated by the radiation ($\rho \propto a^{-4}$ and $H \propto a^{-2}$),
the $e$-folding number given by Eq.~(\ref{eq:deltaN1}) is rewritten as
\begin{eqnarray}
N &\! = \!& \ln \left( {a (t_{\rm reh}) \over a (t_{\rm inf})} \right) 
+ \ln \left( {a(t_{\rm c}) \over a(t_{\rm reh}) } \right) \nonumber\\
&\! = \!&
- {1 \over 6} \ln \left( {\Gamma \over H(t_{\rm inf})} \right)
+{1 \over 2} \ln \left( {H(t_{\rm inf}) \over H(t_{\rm c})} \right) ~,
\end{eqnarray}
where we have used $H(t_{\rm reh}) = \Gamma$.
The fluctuation of $\sigma$ induces the modulated reheating
and hence the fluctuation of the $e$-folding number is given by
\begin{eqnarray}
\delta N = - {1 \over 6} {\delta \Gamma (\sigma) \over \Gamma (\sigma)}
= - {1 \over 6} {\Gamma' \over \Gamma}\delta \sigma~,
\end{eqnarray}
where $\Gamma'(\sigma) \equiv d\Gamma(\sigma) / d\sigma$.
In terms of the reheating temperature $T_R \propto \Gamma^{1/2}$,
the above expression can be rewritten as
\begin{eqnarray}
\delta N = - {1 \over 3}{\delta T_R \over T_R}~.
\label{eq:deltaN2}
\end{eqnarray}
Based on $\delta N$
formalism~\cite{Starobinsky:1986fxa,Sasaki:1995aw,Sasaki:1998ug,
Wands:2000dp,Rigopoulos:2003ak,Lyth:2004gb,Lyth:2005fi},
the curvature perturbation on the uniform energy density hypersurface,
$\zeta$, on super-horizon scales 
is given by the perturbation of the $e$-folding number as
$\zeta \approx \delta N$ and hence
we find that in the modulated reheating scenario
the curvature perturbation can be generated due to the fluctuation
of the decay rate of the inflaton.
Up to the second order,
we have
\begin{eqnarray}
\zeta \approx \delta N = - {1 \over 6}{\Gamma ' \over \Gamma} 
\left[ \delta \sigma
+{1 \over 2} \left( {\Gamma'' \over \Gamma'} - {\Gamma' \over \Gamma}
\right) \delta \sigma^2 \right]~.
\end{eqnarray}
Hence the power spectrum and the non-linearity parameter $f_{\rm NL}$ defined as
\begin{eqnarray}
\langle \zeta ({\bf k}) \zeta ({\bf k}')\rangle 
&\equiv& (2\pi)^3 {2\pi^2 \over k^3} {\cal P}(k) \delta^{(3)}({\bf k} + {\bf k}')~,\\
\zeta &=& \zeta_{\rm lin} + {3 \over 5}f_{\rm NL}\zeta_{\rm lin}^2~,
\end{eqnarray}
are respectively given by
\begin{eqnarray}
{\cal P}(k) &=& \left( {1 \over 6}{\Gamma' \over \Gamma} \right)^2
\left( {H_{\rm inf} \over 2\pi}\right)^2~, \\
f_{\rm NL} &=& 5\left( 1 - {\Gamma \Gamma'' \over \Gamma'^2}\right)~.
\end{eqnarray}

As an example, let us consider an interaction term in  Lagrangian as
$\mathcal{L}_{\rm int} \supset -g(\sigma) \phi \bar{\psi} \psi$ where
$g(\sigma)$ is a coupling constant and $\psi$ is a light fermion which
interacts with the standard particles  and its energy finally goes to
thermal bath.  Then we obtain the decay rate of the inflation $\Gamma$
as
\begin{eqnarray}
\Gamma (\sigma) \simeq \frac{g(\sigma)^2}{8\pi}m_{\phi} ~, 
\end{eqnarray}
where $m_\phi$ is the mass of the inflaton.  Assuming a
$\sigma$-dependence of $g$
as
\begin{eqnarray}
g(\sigma) = g \left[ 1+ \lambda \left({\sigma \over M_{\rm cut}}\right)^2 \right]~,
\end{eqnarray}
where $\lambda ( \sim {\cal O}(1))$ is some constant parameter   and
$M_{\rm cut}$ is a cutoff scale of the effective
theory~\cite{Ackerman:2005,Podolsky:2006,Kohri:2009ac,Kawasaki:2009hp},
the non-linearity parameter $f_{\rm NL}$ is given by
\begin{eqnarray}
f_{\rm NL} \simeq   - 10^2 \lambda^{-1} \left( {\sigma / M_{\rm cut} \over 0.1}\right)^{-2} ~,
\end{eqnarray}
where we have assumed that $\sigma / M_{\rm cut} \ll 1$.  The power
spectrum is also obtained as
\begin{eqnarray}
\mathcal{P}( k ) &\simeq& 10^{-10} \times \nonumber\\
&&\left( {H_{\rm inf} \over 10^{13} {\rm GeV}}\right)^2
\left( { M_{\rm cut} \over 10^{16} {\rm GeV} } \right)^{-2} \lambda^2
\left( { \sigma / M_{\rm cut} \over 0.1} \right)^2~, \nonumber \\
\end{eqnarray}
which can be marginally comparable to the observational value
$\mathcal{P}_{\rm obs}( k ) \simeq  3\times
10^{-10}$~\cite{Komatsu:2010fb} or subdominant for normal scales of
$M_{\rm cut}$ ($\sim 10^{16}$ GeV -- $10^{18}$ GeV)\footnote{
In the subdominant case,
we have to consider a dominant component of the
power spectrum of the curvature perturbations, for example, the inflaton
fluctuation. In such case (so-called ``mixed'' case), the expression of
the non-linearity parameter $f_{\rm NL}$ is changed,
but we can also expect the large non-Gaussianity~\cite{Takahashi:2009cx}.}.

 Hence, in the
modulated reheating scenario we can easily find that  the large
non-Gaussianity ($f_{\rm NL} \sim {\cal O}(10)$) can be generated even
for the high energy inflation with the large $r = 16 \epsilon \sim 0.1
(H_{\rm inf}/10^{13}{\rm GeV})^{2}$ when the model parameters are
appropriately chosen.

\section{Affleck-Dine mechanism}

AD mechanism \cite{Affleck:1984fy,Dine:1995kz} has been known as one
of the powerful  candidates for the successful baryogenesis
mechanism. It can be realized by taking advantage of a flat direction
along which scalar potential vanishes in the global SUSY limit. 
Hereafter we call the complex scalar field that parameterizes the flat direction as AD field $\Phi$ and 
assume that it carries non-zero baryon charge $\beta$. 

Though the scalar potential for the AD field vanishes in the global
SUSY limit,  it is lifted by non-renormalizable terms, the
SUSY-breaking effect and some other effects.  Let us consider a
non-renormalizable superpotential for the AD field given by
\begin{equation}
W_{\rm nr}=\frac{\Phi^{n+3}}{(n+3)M_*^n}, 
\end{equation}
where $M_*$ is the cut-off scale and the positive integer $n$ depends on the flat direction. 
Including the SUSY breaking effect, the induced scalar potential reads
\begin{eqnarray}
V=V_{\not{\rm S}}+ {\left| \Phi \right|^{2n+4} \over
M_*^{2n}}+\left({a_B m_{3/2} \over M_*^n}  \Phi^{n+3} + {\rm h.c.}
\right)~, 
\label{eq:threeTerms}
\end{eqnarray}
where $m_{3/2}$ is a gravitino mass and $a_B$ is a complex numerical factor whose amplitude is of order of unity.
$V_{\not{\rm S}}$ is the soft SUSY breaking effect that depends on the SUSY breaking mechanism. 
The second term is the $F$-term that comes from non-renormalizable operator $W_{\rm nr}$. 
The last term represents the interaction between non-renormalizable operator and the SUSY breaking sector coming from 
supergravity effect, which breaks the $U(1)$ baryon symmetry and is called as the $A$-term. 

During and after inflation, the AD field acquires the Hubble induced mass from the interaction between 
the AD field and the inflaton through the supergravity effect, which can be negative, 
\begin{eqnarray}
V_{\rm H}=-c_{\rm H}H^2|\Phi|^2, 
\end{eqnarray}
where 
$c_{\rm H}$ is a positive numerical factor of order of unity. 
\footnote{In the case where $c_{\rm H}$ is negative, the AD field falls into the origin quickly and does not affect cosmology. }

We must note that even before reheating there can be thermal plasma just after the end of inflation  
as a subdominant component of the universe. Its temperature can be expressed as \cite{kolbturner}
\begin{equation}
T\simeq \left(H M_G T_R^2\right)^{1/4} ,  \label{temposc} 
\end{equation}
where $M_G$ is the reduced Planck scale.
Since the AD field can have interaction with the thermal plasma directly or indirectly, 
it acquires a thermal potential \cite{Anisimov},
\begin{equation}
V_{\rm thermal}\sim \left\{
\begin{array}{ll}
h^2 T^2 |\Phi|^2 & \quad (h|\Phi| \ll T), \\
\alpha_g^2 T^4\log\left(\dfrac{h^2|\Phi|^2}{T^2}\right) & \quad (h |\Phi| \gg T), 
\end{array} \right. \label{thermalpot}
\end{equation}
in addition to above terms. 
Here $h$ is the Yukawa or the gauge coupling constant of the AD field,
and  $\alpha_g \equiv g^2/4\pi$ represents the gauge coupling
constant.  The upper term is the thermal mass  from the thermal
plasma. On the other hand,  the lower one  (called the thermal
logarithmic term) represents the two-loop finite temperature effects
coming from the running of the gauge coupling  with the non-zero value
of the AD field. \footnote{The signature of the thermal logarithmic
term can be both  positive or negative depending on the flat
direction. For example,  the $LH_u$ flat direction receives a positive
thermal logarithmic term  \cite{Fujii:2001zr}. Here we consider only
positive thermal logarithmic terms.  The case with negative thermal
logarithmic term is discussed in Ref. \cite{Kasuya:2003yr} in a
different context.}

Thus, the AD field evolves with the effective potential, 
\begin{equation}
V_{\rm eff}=V+V_{\rm H}+V_{\rm thermal}. 
\end{equation}
During and after inflation, when the Hubble parameter $H$ is
sufficiently large, the AD field settles down to the time-dependent
potential minimum,
\begin{equation}
|\Phi | \simeq (H M_*^n)^{1/(n+1)},  
\end{equation}
and traces its evolution.  Note that there can be several
non-renormalizable operators for the AD field but only the one with
the smallest $n$ determines  the dynamics of the AD field.  Thus
hereafter we consider only smaller $n$ ($n\leq 3$).

Let us consider the evolution of the AD field further.  As the Hubble
parameter decreases, the Hubble induced mass also gets small.   Then,
when $H_{\rm osc}^2 \simeq |V_{\rm eff}^{\prime\prime}| $,  the AD
field (more precisely its radial component) starts to  oscillate
around the origin.  Here the dash denotes the derivative with respect
to $\phi\equiv \sqrt{2} |\Phi|$, and hereafter the subscript ``osc''
indicates that  the parameter or the variable is evaluated at the
onset of the AD field oscillation.

At the onset of the oscillation, the AD field acquires an angular momentum due to the $A$-term, which represents the 
baryon asymmetry of the Universe $n_B$, 
\begin{equation}
n_B (t_{\rm osc})\simeq \beta m_{3/2} (H_{\rm osc}M_*^n)^{2/(n+1)}\sin (n \theta_{\rm inf} + \alpha), 
\end{equation}
where $\theta_{\rm inf}$ and $\alpha$ are the phases of $\Phi$ during
inflation and the constant $a_B$ in the third term of
Eq.~(\ref{eq:threeTerms}), respectively.  Just after the onset of the
AD field oscillation, $a^3 n_B$ is conserved since the $CP$-violating $A$-term comes
to ineffective quickly.  This is because the AD field value continues
decreasing with time during the field oscillation due to the cosmic
expansion.  Since the entropy density decreases as $s \propto a^{-3}$
after the reheating if there is no late-time entropy production,  the
baryon-to-entropy ratio $n_{B}/s$ is conserved.  Thus its present
value is estimated as 
\begin{equation}
\left(\frac{n_B}{s}\right)_0\simeq \frac{\beta m_{3/2} T_R}{M_G^2 H_{\rm osc}^2}(H_{\rm osc} M_*^n)^{2/(n+1)}\sin (n \theta_{\rm inf} + \alpha) . \label{preber}
\end{equation}

Note that there are four scenarios according to which term in the potential drives the AD field oscillation. 
Thus $H_{\rm osc}$ is different from each other  \cite{Fujii:2001zr}. 
The first scenario is that the AD field oscillation is driven by the zero-temperature potential $V_{\not S}$. 
In this case, the Hubble parameter at the onset of the AD field oscillation is 
\begin{equation}
H_{\rm osc} \simeq m_0(\left| \Phi \right|_{\rm osc}), \label{hosc0}
\end{equation}
where $m_0(|\Phi | )\equiv V_{\not S}^{\prime\prime}(| \Phi | )$. 
The second one is that it is driven by the thermal logarithmic term and 
\begin{equation}
H_{\rm osc} \simeq \left(\frac{M_G^{n+1} T_R^{2(n+1)}}{M_*^{2n}}\right)^{1/(n+3)}.   \label{hosclog}
\end{equation}
The third possibility is that it is driven by the thermal mass and 
\begin{equation}
H_{\rm osc} \simeq (h^4 M_GT_R^2)^{1/3}. \label{hosctm}
\end{equation}
When the Yukawa coupling $h$ is rather small, there is a discrepancy in the thermal potential at $|\Phi| \simeq T/h$. 
Thus, if $n<3$, it is possible that AD field start to oscillate immediately after the AD field value becomes  $|\Phi| \simeq T/h$. 
In this case, the Hubble parameter at the onset of the AD field oscillation, 
\begin{equation}
H_{\rm osc} \simeq \left(\frac{M_G^{n+1}T_R^{2(n+1)}}{h^{8(n+1)}M_*^{4n}}\right)^{1/(3-n)}. \label{hoscdel}
\end{equation}
This classification is essential to estimate the baryonic-isocurvature
perturbation as we will see in the next section.

\section{Baryonic-Isocurvature fluctuation}

Let us consider the baryonic-isocurvature fluctuation $S_B$,
which is commonly defined as
\begin{eqnarray}
S_B \equiv {\delta n_B \over n_B} - {\delta s \over s} = { \delta{(n_B / s)} \over n_B / s}~. \label{BIPgen}
\end{eqnarray}
In the case where the baryon-to-entropy ratio depends on the reheating temperature,
the baryonic-isocurvature fluctuation can be also generated in the modulated reheating scenario
\cite{Takahashi:2009dr}.
If we assume $n_B / s \propto T_{\rm R}^p$, we have
\begin{eqnarray}
S_B = { \delta{(n_B / s)} \over n_B / s} = p{\delta T_R \over T_R}~. 
\label{eq:baryiso}
\end{eqnarray}
Hence,  
in the case where the curvature perturbation originates mainly from the modulated reheating mechanism, 
from Eq.~(\ref{eq:deltaN2})  and Eq.~(\ref{eq:baryiso}), we have
\begin{eqnarray}
S_B = - 3p \zeta~.
\end{eqnarray}
The current observational limit for the anti-correlated baryonic-isocurvature fluctuation is 
roughly given by $\left| S_B/\zeta \right|  \lesssim \CO(0.1)$ \cite{Komatsu:2010fb,Kawasaki:2007mb} and hence
it means that the models with $p\gtrsim\CO(1)$ are conflict with current observations.
From  Eq. \eqref{preber}, we find that in general the present baryon-to-entropy ratio generated by the AD mechanism
strongly depends on the reheating temperature. 
Thus it seems that the AD mechanism and the modulated reheating scenario are incompatible.

However, when $n=1$ and $H_{\rm osc}\simeq (M_G T_R^2/M_*)^{1/2}$
(Eq. \eqref{hosclog})  or $n=3$ and $H_{\rm osc}\simeq (h^4M_G
T_R^2)^{1/3} (h\simeq 10^{-5}$) (Eq. \eqref{hosctm}),  the
dependence of the present baryon-to-entropy ratio on the reheating
temperature is canceled.  If such situations are realized,
baryonic-isocurvature fluctuation is not generated from the modulated
reheating.  It is non-trivial whether there is a parameter space in
which the Hubble parameter at the onset of the AD field oscillation
comes to the value above and the baryon-to-entropy ratio can be
explained the present value, $(n_B/s)_0\simeq {\cal O}(10^{-10})$.  In
fact, there {\it is} two sets of parameter spaces in the gravity
mediated SUSY-breaking mechanism \cite{Fujii:2001zr}:  (i) $n=1,
M_*\simeq 10^{22-23}$ GeV, and $T_R \gtrsim 10^7$ GeV, and  (ii) $n=3,
M_*\simeq 10^{16}$ GeV, and $T_R\simeq 10^6$ GeV. \footnote{In
these reheating temperatures, the gravitino problem can be avoidable
\cite{Kawasaki:2008qe}. } In the case
(i) this cut-off scale $M_*$ suggests that small lightest neutrino
mass for the $LH_u$ flat direction \cite{Fujii:2001zr} and  in the
case (ii) it coincides with the GUT scale.  In these cases, the AD mechanism
and the modulated reheating scenario are compatible.

Here we comment on another source of the baryonic-isocurvature perturbation. 
In the absence of the Hubble induced $A$-term, the baryonic-isocurvature perturbation 
is generated from the fluctuation for the angular component of the AD field \cite{Yokoyama:1993gb}. 
The magnitude of the fluctuation is given by 
\begin{equation}
\delta \theta\simeq\frac{H_{\rm inf}}{2\pi|\Phi_{\rm inf}|}\simeq \frac{1}{2\pi}\left(\frac{H_{\rm inf}}{M_*}\right)^{n/(n+1)}, 
\end{equation}
where $H_{\rm inf}$ and $|\Phi_{\rm inf}|$ are the Hubble parameter and the AD field value during inflation. 
Thus, from Eq.~\eqref{preber} and Eq.~\eqref{BIPgen},  we have the net baryonic-isocurvature perturbation as 
\begin{equation}
{\cal S}_B\simeq \frac{n}{2\pi} \cot(n\theta_{\rm inf}+\alpha) \left(\frac{H_{\rm inf}}{M_*}\right)^{n/(n+1)}+ p \frac{\delta T_R}{T_R}. 
\end{equation}

By considering the observational constraint on the uncorrelated
baryonic isocurvature mode ($\left| S_B/\zeta \right|  \lesssim \CO(1)$ 
\cite{Komatsu:2010fb,Kawasaki:2007mb}) 
originated from the fluctuation for the
angular component of the AD field (e.g., see
Ref.~\cite{Kawasaki:2008jy}) we find that the  Hubble parameter during
inflation should be restricted to be (i) $H_{\rm inf} \lesssim
10^{13}$ GeV and (ii) $H_{\rm inf} \lesssim 10^{9}$ GeV, respectively. 
\footnote{Here we assume $\cot (n\theta_{\rm inf}+\alpha) \simeq \CO(1)-\CO(0.1)$. 
This is valid when there is no fine-tuning for the initial phase of the AD field. }
Thus in the case (i) the chaotic inflation with $H_{\rm inf}
\sim 10^{13}$ GeV is still allowed by the observation in terms of the
baryonic-isocurvature fluctuation.   In such case, we can expect
the large tensor-to-scalar ratio at the detectable level in the future
experiments, for examples, such as  Planck satellite, PolarBeaR or
LiteBIRD satellite.

\section{Conclusion}

Recently, the primordial non-Gaussianity and the tensor fluctuations
have been focus of attention to provide the information about
the physics of the early Universe, especially, the origin of the
observed cosmological perturbations.   There would be a lot of the
scenarios generating large non-Gaussianity, and  the ones generating
large tensor-to-scalar ratio, respectively  at the detectable level in
the future experiments independently of each other. 

On the other hand, however, the modulated reheating scenario is quite
attractive since it can produce the large non-Gaussianity, and would
simultaneously predict the large tensor to scalar ratio when the
inflation scale were large like the case of chaotic inflation.

Of course, we need to check the consistencies of
the modulated reheating scenario as a mechanism of generating primordial
curvature fluctuations with other phenomena in the early Universe.
In this paper, we focused on the baryon asymmetry in the Universe. 
In particular, we consider the one of the most promising candidates for baryogenesis, the AD mechanism. 

We have shown that the modulated reheating scenario is consistent with
AD baryogenesis in some sets of model parameters, which can
be motivated from the physics at the high-energy scales,  even if the
observational constraint for the isocurvature fluctuation and the
gravitino problem is imposed on the prediction.  This conclusion is
not changed even if we consider the uncorrelated baryonic-isocurvature
mode originated from the fluctuation of the angular component  in the
AD field. Therefore we will be able to discern this model from others
in principle when we detect a large $f_{\rm NL}$ and a large $r$ in
near future.

\section*{Acknowledgments}

The authors thank the Yukawa Institute for Theoretical Physics at
Kyoto University, where this work was discussed during the
YITP-W-09-05 on "The non-Gaussian universe" and also the YITP-T-10-01 on "Gravity and
Cosmology 2010". We thank Toyokazu Sekiguchi, Fuminobu Takahashi, Masahide Yamaguchi 
and Jun'ichi Yokoyama for discussion.  The work
of S.Y. is supported in part by the Grant-in-Aid for the Global COE
Program ``Quest for Fundamental Principles in the Universe: from
Particles to the Solar System and the Cosmos'' from 
the Ministry of Education, Culture, Sports, Science and Technology (MEXT), Japan
and Grant-in-Aid for Scientific Research No.~22340056. 
K. Kohri was partly supported by the Center for the Promotion of
Integrated Sciences (CPIS) of Sokendai, and Grant-in-Aid for
Scientific Research on Priority Areas No.~18071001,  Scientific
Research (A) No.~22244030 and  Innovative Areas No.~21111006. 
The work of K. Kamada was partly supported by JSPS through research fellowships. 
This work was also supported in part by Global COE Program ``Global Center
of Excellence for Physical Sciences Frontier'', MEXT, Japan.

%%%%%%%%%%%%%%%%%%%%%%%%%%%%%%%%%%%%%%%%%%%%%%%%%%%%%%%%%%%%%%%%%%%%%%


\begin{thebibliography}{999}
%%%%%%%%%%%%%%%%%%%%%%%%%%%%%%%%%%%%%%%%%%%%%%%%%%%%%%%%%%%%%


%%%%%%%%%%%%%%%%%%%%%%%%%%%%%%%%%%%%%%%%%%%%%%%%%%%%%%%%%%%%%




\bibitem{RDcond}

M.~Kawasaki, K.~Kohri and N.~Sugiyama,
  %``Cosmological Constraints on Late-time Entropy Production,''
  Phys.\ Rev.\ Lett.\  {\bf 82}, 4168 (1999);
%  [arXiv:astro-ph/9811437].
  %%CITATION = PRLTA,82,4168;%%
  %``MeV-scale reheating temperature and thermalization of neutrino
  %background,''
  Phys.\ Rev.\  D {\bf 62}, 023506 (2000);
%  [arXiv:astro-ph/0002127].
  %%CITATION = PHRVA,D62,023506;%%
G.~F.~Giudice, E.~W.~Kolb and A.~Riotto,
  %``Largest temperature of the radiation era and its cosmological
  %implications,''
  Phys.\ Rev.\  D {\bf 64}, 023508 (2001);
%  [arXiv:hep-ph/0005123].
  %%CITATION = PHRVA,D64,023508;%%
S.~Hannestad,
  %``What is the lowest possible reheating temperature?,''
  Phys.\ Rev.\  D {\bf 70}, 043506 (2004);
%  [arXiv:astro-ph/0403291].
  %%CITATION = PHRVA,D70,043506;%%
K.~Ichikawa, M.~Kawasaki and F.~Takahashi,
  %``The oscillation effects on thermalization of the neutrinos in the universe
  %with low reheating temperature,''
  Phys.\ Rev.\  D {\bf 72}, 043522 (2005). 
%  [arXiv:astro-ph/0505395].
  %%CITATION = PHRVA,D72,043522;%%


\bibitem{textbook}
%\bibitem{Linde:2005ht}
  A.~D.~Linde,
  ``Particle Physics and Inflationary Cosmology,''
   Chur, Switzerland: Harwood (1990) 362 p. ;
%  arXiv:hep-th/0503203;
  %%CITATION = HEP-TH/0503203;%%
%\bibitem{Lyth:1998xn}
  D.~H.~Lyth and A.~Riotto,
  %``Particle physics models of inflation and the cosmological density
  %perturbation,''
  Phys.\ Rept.\  {\bf 314}, 1 (1999);
%  [arXiv:hep-ph/9807278];
  %%CITATION = PRPLC,314,1;%%
%\bibitem{Liddle:2000cg}
  A.~R.~Liddle and D.~H.~Lyth
  ``Cosmological inflation and large-scale structure,''
  {\it   Cambridge, UK: Univ. Pr. (2000) 400 p} ;
  %%CITATION = ISBN-13-9780521828499;%%
%\bibitem{Mukhanov:2005sc}
  V.~Mukhanov,
  ``Physical foundations of cosmology,''
%\href{http://www.slac.stanford.edu/spires/find/hep/www?irn=6927394}{SPIRES entry}
{\it  Cambridge, UK: Univ. Pr. (2005) 421 p};
%\bibitem{Weinberg:2008zzc}
  S.~Weinberg,
  ``Cosmology,''
%\href{http://www.slac.stanford.edu/spires/find/hep/www?irn=7886489}{SPIRES entry}
{\it  Oxford, UK: Oxford Univ. Pr. (2008) 593 p};
%\bibitem{Lyth:2009zz}
  D.~H.~Lyth and A.~R.~Liddle,
  ``The primordial density perturbation: Cosmology, inflation and the origin of
  structure,''
%\href{http://www.slac.stanford.edu/spires/find/hep/www?irn=8640165}{SPIRES entry}
{\it  Cambridge, UK: Cambridge Univ. Pr. (2009) 497 p};
%\bibitem{Mazumdar:2010sa}
  A.~Mazumdar and J.~Rocher,
  %``Particle physics models of inflation and curvaton scenarios,''
  arXiv:1001.0993 [hep-ph].
  %%CITATION = ARXIV:1001.0993;%%


\bibitem{starobinsky:1985}
A. A. Starobinsky, Pisma Zh. Eksp. Teor. Fiz. {\bf 42}, 124 (1985)
[JETP Lett. {\bf 42}, 152 (1985)].

\bibitem{Alabidi:2006hg}
  L.~Alabidi,
  %``Non-gaussianity for a two component hybrid model of inflation,''
  JCAP {\bf 0610}, 015 (2006)
  [arXiv:astro-ph/0604611].
  %%CITATION = JCAPA,0610,015;%%

\bibitem{curvatonNG}
%\bibitem{LUW20003}
D. H. Lyth, C. Ungarelli, and D. Wands,
  %``The primordial density perturbation in the curvaton scenario,''
  Phys. Rev.  D {\bf 67}, 023503 (2003);
  %[arXiv:astro-ph/0208055].
  %%CITATION = PHRVA,D67,023503;%%
%\bibitem{Lyth:2003ip}
  D.~H.~Lyth and D.~Wands,
  %``The CDM isocurvature perturbation in the curvaton scenario,''
  Phys.\ Rev.\  D {\bf 68}, 103516 (2003);
%  [arXiv:astro-ph/0306500].
  %%CITATION = PHRVA,D68,103516;%%
%\bibitem{Bartolo:2003jx}
  N.~Bartolo, S.~Matarrese and A.~Riotto,
  %``On non-Gaussianity in the curvaton scenario,''
  Phys.\ Rev.\  D {\bf 69}, 043503 (2004);
%  [arXiv:hep-ph/0309033].
  %%CITATION = PHRVA,D69,043503;%%
%\bibitem{Enqvist:2005pg}
  K.~Enqvist and S.~Nurmi,
  %``Non-gaussianity in curvaton models with nearly quadratic potential,''
  JCAP {\bf 0510}, 013 (2005);
%  [arXiv:astro-ph/0508573].
  %%CITATION = JCAPA,0510,013;%%
%\bibitem{Malik:2006pm}
  K.~A.~Malik and D.~H.~Lyth,
  %``A numerical study of non-gaussianity in the curvaton scenario,''
  JCAP {\bf 0609}, 008 (2006);
%  [arXiv:astro-ph/0604387].
  %%CITATION = JCAPA,0609,008;%%
%\bibitem{Sasaki:2006kq}
  M.~Sasaki, J.~Valiviita and D.~Wands,
  %``Non-gaussianity of the primordial perturbation in the curvaton model,''
  Phys.\ Rev.\  D {\bf 74}, 103003 (2006);
%  [arXiv:astro-ph/0607627].
  %%CITATION = PHRVA,D74,103003;%%
%\bibitem{Huang:2008ze}
  Q.~G.~Huang,
  %``Large Non-Gaussianity Implication for Curvaton Scenario,''
  Phys.\ Lett.\  B {\bf 669}, 260 (2008);
%  [arXiv:0801.0467 [hep-th]].
  %%CITATION = PHLTA,B669,260;%%
%\bibitem{Ichikawa:2008iq}
  K.~Ichikawa, T.~Suyama, T.~Takahashi and M.~Yamaguchi,
  %``Non-Gaussianity, Spectral Index and Tensor Modes in Mixed Inflaton and
  %Curvaton Models,''
  Phys.\ Rev.\  D {\bf 78}, 023513 (2008);
%  [arXiv:0802.4138 [astro-ph]].
  %%CITATION = PHRVA,D78,023513;%%
%\bibitem{Huang:2008bg}
  Q.~G.~Huang and Y.~Wang,
  %``Curvaton Dynamics and the Non-Linearity Parameters in Curvaton Model,''
  JCAP {\bf 0809}, 025 (2008);
%  [arXiv:0808.1168 [hep-th]].
  %%CITATION = JCAPA,0809,025;%%
%\bibitem{Enqvist:2008gk}
  K.~Enqvist and T.~Takahashi,
  %``Signatures of Non-Gaussianity in the Curvaton Model,''
  JCAP {\bf 0809}, 012 (2008);
%  [arXiv:0807.3069 [astro-ph]].
  %%CITATION = JCAPA,0809,012;%%
%\bibitem{Huang:2008zj}
  Q.~G.~Huang,
  %``Curvaton with Polynomial Potential,''
  JCAP {\bf 0811}, 005 (2008);
%  [arXiv:0808.1793 [hep-th]].
  %%CITATION = JCAPA,0811,005;%%
%\bibitem{Moroi:2008nn}
T.~Moroi and T.~Takahashi,
  %``Non-Gaussianity and Baryonic Isocurvature Fluctuations in the Curvaton
  %Scenario,''
  Phys.\ Lett.\  B {\bf 671}, 339 (2009);
%  [arXiv:0810.0189 [hep-ph]].
  %%CITATION = PHLTA,B671,339;%
%\bibitem{Kawasaki:2008mc}
  M.~Kawasaki, K.~Nakayama and F.~Takahashi,
  %``Hilltop Non-Gaussianity,''
  JCAP {\bf 0901}, 026 (2009);
%  [arXiv:0810.1585 [hep-ph]].
  %%CITATION = JCAPA,0901,026;%%
%\bibitem{Chingangbam:2009xi}
  P.~Chingangbam and Q.~G.~Huang,
  %``The Curvature Perturbation in the Axion-type Curvaton Model,''
  JCAP {\bf 0904}, 031 (2009);
%  [arXiv:0902.2619 [astro-ph.CO]].
  %%CITATION = JCAPA,0904,031;%%
%\bibitem{Enqvist:2009zf}
  K.~Enqvist, S.~Nurmi, G.~Rigopoulos, O.~Taanila and T.~Takahashi,
  %``The Subdominant Curvaton,''
  JCAP {\bf 0911}, 003 (2009);
%  [arXiv:0906.3126 [astro-ph.CO]].
  %%CITATION = JCAPA,0911,003;%%
%\bibitem{Enqvist:2009eq}
  K.~Enqvist and T.~Takahashi,
  %``Effect of Background Evolution on the Curvaton Non-Gaussianity,''
  JCAP {\bf 0912}, 001 (2009);
%  [arXiv:0909.5362 [astro-ph.CO]].
  %%CITATION = JCAPA,0912,001;%%
%\bibitem{Enqvist:2009ww}
  K.~Enqvist, S.~Nurmi, O.~Taanila and T.~Takahashi,
  %``Non-Gaussian Fingerprints of Self-Interacting Curvaton,''
  JCAP {\bf 1004}, 009 (2010);
%  [arXiv:0912.4657 [astro-ph.CO]].
  %%CITATION = JCAPA,1004,009;%%
%\bibitem{Chingangbam:2010xn}
  P.~Chingangbam and Q.~G.~Huang,
  %``New features in curvaton model,''
  arXiv:1006.4006 [astro-ph.CO];
  %%CITATION = ARXIV:1006.4006;%%
%\bibitem{Byrnes:2010xd}
  C.~T.~Byrnes, K.~Enqvist and T.~Takahashi,
  %``Scale-dependence of Non-Gaussianity in the Curvaton Model,''
  arXiv:1007.5148 [astro-ph.CO].
  %%CITATION = ARXIV:1007.5148;%%



\bibitem{Bernardeau:2003}
  F.~Bernardeau and J.~P.~Uzan,
  %``Inflationary models inducing non-gaussian metric fluctuations,''
  Phys.\ Rev.\  D {\bf 67} (2003) 121301;
%  [arXiv:astro-ph/0209330].
  %%CITATION = PHRVA,D67,121301;%%
  F.~Bernardeau, L.~Kofman and J.~P.~Uzan,
  %``Modulated fluctuations from hybrid inflation,''
  Phys.\ Rev.\  D {\bf 70} (2004) 083004;
  [arXiv:astro-ph/0403315].
  %%CITATION = PHRVA,D70,083004;%%

\bibitem{Lyth:2005}
 D.~H.~Lyth,
  %``Generating the curvature perturbation at the end of inflation,''        
  JCAP {\bf 0511}, 006 (2005).
%  [arXiv:astro-ph/0510443].                                                 
  %%CITATION = JCAPA,0511,006;%%    

\bibitem{Alabidi:2006wa}
  L.~Alabidi and D.~Lyth,
  %``Curvature perturbation from symmetry breaking the end of inflation,''
  JCAP {\bf 0608}, 006 (2006)
  [arXiv:astro-ph/0604569].
  %%CITATION = JCAPA,0608,006;%%


\bibitem{Lin:2009vj}
  C.~M.~Lin,
  %``Large non-Gaussianity generated at the end of Extended D-term Hybrid
  %Inflation,''
  arXiv:0908.4168 [hep-ph].
  %%CITATION = ARXIV:0908.4168;%%



\bibitem{Huang:2009vk}
  Q.~G.~Huang,
  %``A geometric description of the non-Gaussianity generated at the end of
  %multi-field inflation,''
  JCAP {\bf 0906}, 035 (2009)
  [arXiv:0904.2649 [hep-th]].
  %%CITATION = JCAPA,0906,035;%%

\bibitem{modureh}
 G.~Dvali, A.~Gruzinov and M.~Zaldarriaga,
  %``A new mechanism for generating density perturbations from inflation,''
  Phys.\ Rev.\  D {\bf 69} (2004) 023505;
%  [arXiv:astro-ph/0303591].
  %%CITATION = PHRVA,D69,023505;%%
 L.~Kofman,
  %``Probing string theory with modulated cosmological fluctuations,''
  arXiv:astro-ph/0303614;
  %%CITATION = ASTRO-PH/0303614;%%
 G.~Dvali, A.~Gruzinov and M.~Zaldarriaga,
  %``Cosmological perturbations from inhomogeneous reheating, freezeout, and
  %mass domination,''
  Phys.\ Rev.\  D {\bf 69} (2004) 083505;
%  [arXiv:astro-ph/0305548].
  %%CITATION = PHRVA,D69,083505;%%
  M.~Zaldarriaga,
  %``Non-Gaussianities in models with a varying inflaton decay rate,''
  Phys.\ Rev.\  D {\bf 69} 043508 (2004). 
%  [arXiv:astro-ph/0306006].
  %%CITATION = PHRVA,D69,043508;%%




\bibitem{Mazumdar:2003va}
  A.~Mazumdar,
  %``A model for fluctuating inflaton coupling: (s)neutrino induced  adiabatic
  %perturbations and non-thermal leptogenesis,''
  Phys.\ Rev.\ Lett.\  {\bf 92}, 241301 (2004)
  [arXiv:hep-ph/0306026].
  %%CITATION = PRLTA,92,241301;%%


\bibitem{Ackerman:2005}
  L.~Ackerman, C.~W.~Bauer, M.~L.~Graesser and M.~B.~Wise,
  %``Light scalars and the generation of density perturbations during
  %preheating or inflaton decay,''
  Phys.\ Lett.\  B {\bf 611} (2005) 53.
 % [arXiv:astro-ph/0412007].
  %%CITATION = PHLTA,B611,53;%%

\bibitem{Podolsky:2006}
  D.~I.~Podolsky, G.~N.~Felder, L.~Kofman and M.~Peloso,
  %``Equation of state and beginning of thermalization after preheating,''
  Phys.\ Rev.\  D {\bf 73} (2006) 023501.
%  [arXiv:hep-ph/0507096].
  %%CITATION = PHRVA,D73,023501;%%

\bibitem{Suyama:2007}
  T.~Suyama and M.~Yamaguchi,
  %``Non-Gaussianity in the modulated reheating scenario,''
  Phys.\ Rev.\  D {\bf 77}, 023505 (2008);
%  [arXiv:0709.2545 [astro-ph]]
  %%CITATION = PHRVA,D77,023505;%%
%\bibitem{Ichikawa:2008ne}
  K.~Ichikawa, T.~Suyama, T.~Takahashi and M.~Yamaguchi,
  %``Primordial Curvature Fluctuation and Its Non-Gaussianity in Models with
  %Modulated Reheating,''
  Phys.\ Rev.\  D {\bf 78}, 063545 (2008).
%  [arXiv:0807.3988 [astro-ph]].
  %%CITATION = PHRVA,D78,063545;%%

\bibitem{Kohri:2009ac}
  K.~Kohri, D.~H.~Lyth and C.~A.~Valenzuela-Toledo,
  %``On the generation of a non-gaussian curvature perturbation during
  %preheating,''
  JCAP {\bf 1002}, 023 (2010)
  [arXiv:0904.0793 [hep-ph]].
  %%CITATION = JCAPA,1002,023;%%

\bibitem{otherModpre}
%\bibitem{Battefeld:2007st}
  T.~Battefeld,
  %``Modulated Perturbations from Instant Preheating after new Ekpyrosis,''
  Phys.\ Rev.\  D {\bf 77}, 063503 (2008)
  [arXiv:0710.2540 [hep-th]]; 
  %%CITATION = PHRVA,D77,063503;%%
%\bibitem{Byrnes:2008zz}
  C.~T.~Byrnes,
  %``Constraints on generating the primordial curvature perturbation and
  %non-Gaussianity from instant preheating,''
  JCAP {\bf 0901}, 011 (2009)
  [arXiv:0810.3913 [astro-ph]];
  %%CITATION = JCAPA,0901,011;%%
%\bibitem{Matsuda:2007tr}
  T.~Matsuda,
  %``Cosmological perturbations from inhomogeneous preheating and multi-field
  %trapping,''
  JHEP {\bf 0707}, 035 (2007)
  [arXiv:0707.0543 [hep-th]].
  %%CITATION = JHEPA,0707,035;%%


\bibitem{Matsuda:2009yt}
  T.~Matsuda,
  %``Cosmological perturbations from an inhomogeneous phase transition,''
  arXiv:0902.4283 [hep-ph].
  %%CITATION = ARXIV:0902.4283;%%


\bibitem{Kawasaki:2009hp}
  M.~Kawasaki, T.~Takahashi and S.~Yokoyama,
  %``Density Fluctuations in Thermal Inflation and Non-Gaussianity,''
  JCAP {\bf 0912}, 012 (2009)
  [arXiv:0910.3053 [hep-th]].
  %%CITATION = JCAPA,0912,012;%%



\bibitem{Langlois:2009jp}
  D.~Langlois and L.~Sorbo,
  %``Primordial perturbations and non-Gaussianities from modulated trapping,''
  JCAP {\bf 0908}, 014 (2009)
  [arXiv:0906.1813 [astro-ph.CO]].
  %%CITATION = JCAPA,0908,014;%%


\bibitem{Komatsu:2010fb}
  E.~Komatsu {\it et al.},
  %``Seven-Year Wilkinson Microwave Anisotropy Probe (WMAP) Observations:
  %Cosmological Interpretation,''
  arXiv:1001.4538 [astro-ph.CO].
  %%CITATION = ARXIV:1001.4538;%%

\bibitem{PLANCK_HP}
http://www.rssd.esa.int/index.php?project=Planck


\bibitem{PLANCK_coll}
    [Planck Collaboration],
  %``Planck: The scientific programme,''
  arXiv:astro-ph/0604069.
  %%CITATION = ASTRO-PH/0604069;%%


\bibitem{Komatsu:2009kd}
  E.~Komatsu {\it et al.},
  %``Non-Gaussianity as a Probe of the Physics of the Primordial Universe and
  %the Astrophysics of the Low Redshift Universe,''
  arXiv:0902.4759 [astro-ph.CO].
  %%CITATION = ARXIV:0902.4759;%%

\bibitem{Hazumi:2008zz}
  M.~Hazumi,
  %``Jumping into CMB polarization measurements: A new group at KEK,''
  AIP Conf.\ Proc.\  {\bf 1040}, 78 (2008).
  %%CITATION = APCPC,1040,78;%%

\bibitem{LiteBird}
http://cmbpol.kek.jp/litebird/



%%%%%%%%%%%%%%%%%%%%%%%%%%%%%%%%%%%%%%%%%%%%%%

%\cite{Takahashi:2009dr}
\bibitem{Takahashi:2009dr}
  T.~Takahashi, M.~Yamaguchi, J.~Yokoyama and S.~Yokoyama,
  %``Gravitino Dark Matter and Non-Gaussianity,''
  Phys.\ Lett.\  B {\bf 678}, 15 (2009). 

%\cite{Takahashi:2009cx}
\bibitem{Takahashi:2009cx}
  T.~Takahashi, M.~Yamaguchi and S.~Yokoyama,
  %``Primordial Non-Gaussianity in Models with Dark Matter Isocurvature
  %Fluctuations,''
  Phys.\ Rev.\  D {\bf 80}, 063524 (2009)
 % [arXiv:0907.3052 [astro-ph.CO]].
  %%CITATION = PHRVA,D80,063524;%%


%\cite{Affleck:1984fy}
\bibitem{Affleck:1984fy}
%\bibitem{AD}
  I.~Affleck and M.~Dine,
  %``A New Mechanism For Baryogenesis,''
  Nucl.\ Phys.\  B {\bf 249}, 361 (1985). 
  %%CITATION = NUPHA,B249,361;%%
%\cite{Dine:1995kz}



\bibitem{Dine:1995kz}
  M.~Dine, L.~Randall and S.~D.~Thomas,
  %``Baryogenesis From Flat Directions Of The Supersymmetric Standard Model,''
  Nucl.\ Phys.\  B {\bf 458}, 291 (1996)
  [arXiv:hep-ph/9507453].
  %%CITATION = NUPHA,B458,291;%%

%%%%%%%%%%\subsection{Ref:deltaN}%%%%%%%%%%%%%%%%%%%%%%%%%%%%%%%%%%%%%%%%%%%%%%%%

\bibitem{Starobinsky:1986fxa}
  A.~A.~Starobinsky,
  %``Multicomponent de Sitter (Inflationary) Stages and the Generation of
  %Perturbations,''
  JETP Lett.\  {\bf 42} (1985) 152
  [Pisma Zh.\ Eksp.\ Teor.\ Fiz.\  {\bf 42} (1985) 124]. 
  %%CITATION = ZFPRA,42,124;%%
  
%\cite{Sasaki:1995aw}
\bibitem{Sasaki:1995aw}
M.~Sasaki and E.~D.~Stewart,
  %``A General Analytic Formula For The Spectral Index Of The Density
  %Perturbations Produced During Inflation,''
  Prog.\ Theor.\ Phys.\  {\bf 95}, 71 (1996) 
  [arXiv:astro-ph/9507001]. 
  %%CITATION = ASTRO-PH 9507001;%%  
  
\bibitem{Sasaki:1998ug}
  M.~Sasaki and T.~Tanaka,
  %``Super-horizon scale dynamics of multi-scalar inflation,''
  Prog.\ Theor.\ Phys.\  {\bf 99}, 763 (1998) 
  [arXiv:gr-qc/9801017].
  %%CITATION = GR-QC 9801017;%%  

\bibitem{Wands:2000dp}
  D.~Wands, K.~A.~Malik, D.~H.~Lyth and A.~R.~Liddle,
  %``A new approach to the evolution of cosmological perturbations on large
  %scales,''
  Phys.\ Rev.\  D {\bf 62}, 043527 (2000)
  [arXiv:astro-ph/0003278].
  %%CITATION = PHRVA,D62,043527;%%

\bibitem{Rigopoulos:2003ak}
  G.~I.~Rigopoulos and E.~P.~S.~Shellard,
  %``The Separate Universe Approach and the Evolution of Nonlinear Superhorizon
  %Cosmological Perturbations,''
  Phys.\ Rev.\  D {\bf 68}, 123518 (2003)
  [arXiv:astro-ph/0306620].
  %%CITATION = PHRVA,D68,123518;%%

\bibitem{Lyth:2004gb}
  D.~H.~Lyth, K.~A.~Malik and M.~Sasaki,
  %``A general proof of the conservation of the curvature perturbation,''
  JCAP {\bf 0505}, 004 (2005)
  [arXiv:astro-ph/0411220].
  %%CITATION = JCAPA,0505,004;%%

\bibitem{Lyth:2005fi}
  D.~H.~Lyth and Y.~Rodriguez,
  %``The inflationary prediction for primordial non-gaussianity,''
  Phys.\ Rev.\ Lett.\  {\bf 95}, 121302 (2005)
  [arXiv:astro-ph/0504045].
  %%CITATION = PRLTA,95,121302;%%


%%%%%%%%%%%%%%%%%%%%%%%%%%%%%%%%%%%%%%%%%%%%%%%%%%%%%%%%%%%%%%%%%
  
  \bibitem{kolbturner}
E.W. Kolb and M.S. Turner, {\it The Early Universe}, 
(Westview Press, Boulder, CO, 1990).

%\cite{Anisimov:2000wx}
\bibitem{Anisimov}
  A.~Anisimov and M.~Dine,
  %``Some issues in flat direction baryogenesis,''
  Nucl.\ Phys.\  B {\bf 619}, 729 (2001)
  [arXiv:hep-ph/0008058].
  %%CITATION = NUPHA,B619,729;%%%\cite{Gherghetta:1995dv}
  
\bibitem{Fujii:2001zr}
%\cite{Fujii:2001zr}%\cite{Asaka:2000nb}
%\bibitem{Asaka:2000nb}
  T.~Asaka, M.~Fujii, K.~Hamaguchi and T.~Yanagida,
  %``Affleck-Dine leptogenesis with an ultralight neutrino,''
  Phys.\ Rev.\  D {\bf 62}, 123514 (2000)
  [arXiv:hep-ph/0008041];
  %%CITATION = PHRVA,D62,123514;%%

  M.~Fujii, K.~Hamaguchi and T.~Yanagida,
  %``Reheating-temperature independence of cosmological baryon asymmetry in
  %Affleck-Dine leptogenesis,''
  Phys.\ Rev.\  D {\bf 63}, 123513 (2001)
  [arXiv:hep-ph/0102187]; 
  %%CITATION = PHRVA,D63,123513;%%
  M.~Fujii, master thesis (in Japanese). 
%\cite{Kasuya:2003yr}
\bibitem{Kasuya:2003yr}
  S.~Kasuya, M.~Kawasaki and F.~Takahashi,
  %``Affleck-Dine mechanism with negative thermal logarithmic potential,''
  Phys.\ Rev.\  D {\bf 68}, 023501 (2003)
  [arXiv:hep-ph/0302154].
  %%CITATION = PHRVA,D68,023501;%%
  %\cite{Kawasaki:2007mb}
\bibitem{Kawasaki:2007mb}
  M.~Kawasaki and T.~Sekiguchi,
  %``Cosmological Constraints on Isocurvature and Tensor Perturbations,''
  Prog.\ Theor.\ Phys.\  {\bf 120}, 995 (2008)
  [arXiv:0705.2853 [astro-ph]].
  %%CITATION = PTPKA,120,995;%%

%\cite{Kawasaki:2008qe}
\bibitem{Kawasaki:2008qe}
 M.~Kawasaki, K.~Kohri and T.~Moroi,
 %``Big-bang nucleosynthesis and hadronic decay of long-lived massive
 %particles,''
 Phys.\ Rev.\  D {\bf 71}, 083502 (2005);
 %[arXiv:astro-ph/0408426].
 %%CITATION = PHRVA,D71,083502;%%
 %``Hadronic decay of late-decaying particles and big-bang nucleosynthesis,''
 Phys.\ Lett.\  B {\bf 625}, 7 (2005); 
 %[arXiv:astro-ph/0402490].
 %%CITATION = PHLTA,B625,7;%%
  M.~Kawasaki, K.~Kohri, T.~Moroi and A.~Yotsuyanagi,
  %``Big-Bang Nucleosynthesis and Gravitino,''
  Phys.\ Rev.\  D {\bf 78}, 065011 (2008)
  [arXiv:0804.3745 [hep-ph]].
  %%CITATION = PHRVA,D78,065011;%%



%\cite{Yokoyama:1993gb}
\bibitem{Yokoyama:1993gb}
  J.~Yokoyama,
  %``Formation of baryon number fluctuation in supersymmetric inflationary
  %cosmology,''
  Astropart.\ Phys.\  {\bf 2}, 291 (1994); 
  %%CITATION = APHYE,2,291;%%%\cite{Senami:2007wz}
  M.~Senami and T.~Takayama,
  %``Affleck-Dine leptogenesis via multiscalar evolution in a supersymmetric
  %seesaw model,''
  JCAP {\bf 0711}, 015 (2007)
  [arXiv:0708.2238 [hep-ph]]; 
  %%CITATION = JCAPA,0711,015;%%
  K.~Kamada and J.~Yokoyama,
  %``Affleck Dine leptogenesis via multiple flat directions,''
  Phys.\ Rev.\  D {\bf 78}, 043502 (2008)
  [arXiv:0803.3146 [hep-ph]]; 
  %%CITATION = PHRVA,D78,043502;%%
  S.~Kasuya, M.~Kawasaki and F.~Takahashi,
  %``Isocurvature fluctuations in Affleck-Dine mechanism and constraints on
  %inflation models,''
  JCAP {\bf 0810}, 017 (2008)
  [arXiv:0805.4245 [hep-ph]].
  %%CITATION = JCAPA,0810,017;%%



%%%%%%%%%%%%%%%%%%%%%%%%%%%%%%%%%%%%%%%%%%%%%

%\cite{Kawasaki:2008jy}
\bibitem{Kawasaki:2008jy}
  M.~Kawasaki, K.~Nakayama and F.~Takahashi,
  %``Non-Gaussianity from Baryon Asymmetry,''
  JCAP {\bf 0901}, 002 (2009)
  [arXiv:0809.2242 [hep-ph]].
  %%CITATION = JCAPA,0901,002;%%



  \end{thebibliography}
\end{document}